# A Multifaceted Panel Data Gravity Model Analysis of Peru's Foreign Trade

Xu Wang[1], Ryan P. Badman[2]
[1]Shanghai, China
[2]Cornell University, 109 Clark Hall, Ithaca, NY

**Abstract:** Peru's abundant natural resources and friendly trade policies have made the country a major economic player in both South America and the global community. Consequently, exports are playing an increasingly important role in Peru's national economy. Indeed, growing from 13.1% as of 1994, exports now contribute approximately 21% of the GDP of Peru as of 2015. Given Peru's growing global influence, the time is ripe for a thorough analysis of the most important factors governing its export performance. Thus, within the framework of the augmented gravity model of trade, this paper examines Peru's export performance and attempts to identify the dominant economic factors that should be further developed to increase the value of exports. The analysis was conducted from three different aspects: (1) general economic parameters' effect on Peru's export value, (2) more specific analysis into a major specific trade good, copper, and (3) the impact that regional trade agreements have had on Peru's export performance. Our panel data analysis results for each dataset revealed interesting economic trends and were consistent with the theoretical expectations of the gravity model: namely positive coefficients for economic size and negative coefficients for distance. This report's results can be a reference for the proper direction of Peruvian economic policy so as to enhance economic growth in a sustainable direction.

## 1. Introduction

Between the mid-1970s and 1990, negative growth and hyperinflation shadowed Peru's economy. In the beginning of the 1980s, profoundly affected by internal military conflict, the economy of Peru faced challenges in almost every sector. In 1980, the production of agriculture, mining and fishing shrank 12%, 2.8% and 2.5% respectively compared with that of 1979 [1]. Therefore, a fundamental change in development strategy was urged in Peru. Priority was allocated to the development of agriculture, mining industry and the petrol industry, encouraging privatization by both domestic and foreign capital since then. As a result, a 5.6% GDP growth was achieved in 1982 [2]. However, the economic crisis in Latin America took place in 1983 and hit the economic development of Peru directly, immediately followed by intensified internal military conflict and disagreements with the International Monetary Fund ("IMF") [3]. The general Peruvian economic structure was seriously damaged when extreme incentive plans were carried out by the government of Garcia so as to stimulate the domestic demand. The economic and political chaos in Peru also impeded the influx of capital from international sources in the late 1980s. A reform took place in 1990 where the government agreed to resume normal relations with the IMF, World Bank and IDB [3]. Since the reform in the early 1990s, a substantial increase has been witnessed in the magnitude of investment and trade. Between 1994 and 1998, an average annual growth rate at 8.5% took place in the area of merchandise trade while GDP grew at an average of 5.6% per year from 1993-1999 [4].

Since 2000, the improvement in Peru's economy was mainly attributed to the development of trade. Peru has continued to liberalize and expand its trade strategy mostly through unilateral initiatives, by capping its tariff levels and by introducing measures to facilitate its trade [5]. These policy changes in return built up a favorable economic environment that set up a more stable economic structure, with exports of goods and services expanding at an annual rate of 8.3% from 2000-2006 [5]. Since 2007, Peru's pursuit of and strong contributions to regional trade agreements ("RTAs") has led to an exceptional economic performance [6]. Specifically, Peru has been participating actively in Mercado Comun del Sur ("MERCOSUR") and Andean Community of Nations ("CAN"). MERCOSUR is one of the most influential commercial blocks that have emerged over the last few decades as a customs union in Latin America. Its original members are Brazil, Argentina, Paraguay, and Uruguay. Other associated members are Bolivia (1996), Chile (1996), Peru (2003), Colombia (2004), and Ecuador (2004). Since its creation in 1991, commercial links amongst them have been reinforced and new contacts with other countries have been established. CAN is another important economic organization in Latin and South America, consisting of the member states of Bolivia, Colombia, Ecuador, and Peru. The trade bloc, centered in Lima, Peru, was founded in 1969 and named the Andean Pact up to 1996.

The economic health improvements in Peru were notably demonstrated recently by robust growth in real GDP, with an annual average growth rate of almost 7% during 2007-2012 [6], low inflation at approximately 3.3% from 2008-2014 [6], a balanced fiscal position, reduced debt and improved external accounts. Hence, it seems that the strong performance of the export system has had a profound impact on the economic growth of Peru.

The major components of Peru's export are either raw materials or semi-processed goods in the areas of mining, agricultural products and animal feed, and fishing goods (see Graph 1.1). As a result, the Peruvian economy is vulnerable to large fluctuations as trade depends largely on the global prices of a handful of products. In recent years, Peru has significantly raised its dependence on copper and gold, which in 2012 made up a whopping 80% of its mineral exports (mostly sent to the Chinese market) and 10% of the GDP [6]. In this case, certain products have had an extensive impact on the general performance of export in Peru.

Broadly, the benefits of increasing trade exports can be generally summarized as follows: Exports will expose domestic industries to international markets. The development of manufacturing industries increases the international competitiveness of the products and therefore leads to an increase in potential market. Economies are expected to scale up and grow production capacity with the increase in potential market size. Exposure to international markets can also intensify intra-industry and inter-industry competition, spurring innovation and strengthening the comparative advantage of technology and knowledge in the domestic economy. Another benefit that comes along with the expansion of exporting is increased volume of foreign direct investment ("FDI") which is essential to the economic development of a country. Thus, a growth in the general economy can be reasonably expected following better export performance.

As Peru is heavily dependent on exports, identifying the key variables affecting the export performance of Peru is of great necessity. A focus on these factors is expected to aid the creation of policies to improve economic growth in a sustainable manner. A general model can help to pinpoint the most influential factors. However, a model that is too general may overlook specific real world factors which are influential to the export performance of a specific industry. Therefore, detailed modeling on one or more specific major industries is useful as a supplement. Considering the increasingly important role of the copper export of Peru, a gravity model analysis of the copper industry was chosen to identify the unique aspects that might have been masked by the analysis on the general performance of Peru's exports. On the other hand, given the fact that the top of the list of Peru's most lucrative trade partners is dominated by countries in Europe, North America and Eastern Asian (see Graph 1.1), the importance of the participation in CAN and MERCOSUR is expected to be concealed in a broad scale analysis given that the Peru export volume is dominated by non-regional partners. In order to reveal the hidden regional effects, an analysis was conducted, focusing only on Peru's export toward the members and associate members of CAN and MERCOSUR. Due to the lack of data for CAN, this study focused on the impact of the membership in MERCOSUR.

Literature discussing the application of the gravity model to the modern case of the Peru is limited. In 2004, the World Trade Organization (WTO) conducted a study on Peru's foreign trade using the gravity model. Other studies using standard panel data models or other models for Peru focused only on specific trade sectors including tourism [7], sugar trade [8] and fishmeal trade [9]. None of these studies on specific goods analyzed the copper industry and the 2004 WTO general study is now outdated. Thus, this study is meant to fill the niche in the current literature. In addition, this paper verified that in some cases, certain regional effects will be masked in the gravity model when analyzing from a high level economic perspective. For example: if one export good, such as copper in this study, has massive influence on the trade of the country of interest, or if a small number of trade partners dominate the export destinations. This unusual situation is absent in most gravity model studies such as the Philippines [10] and Uganda [11].

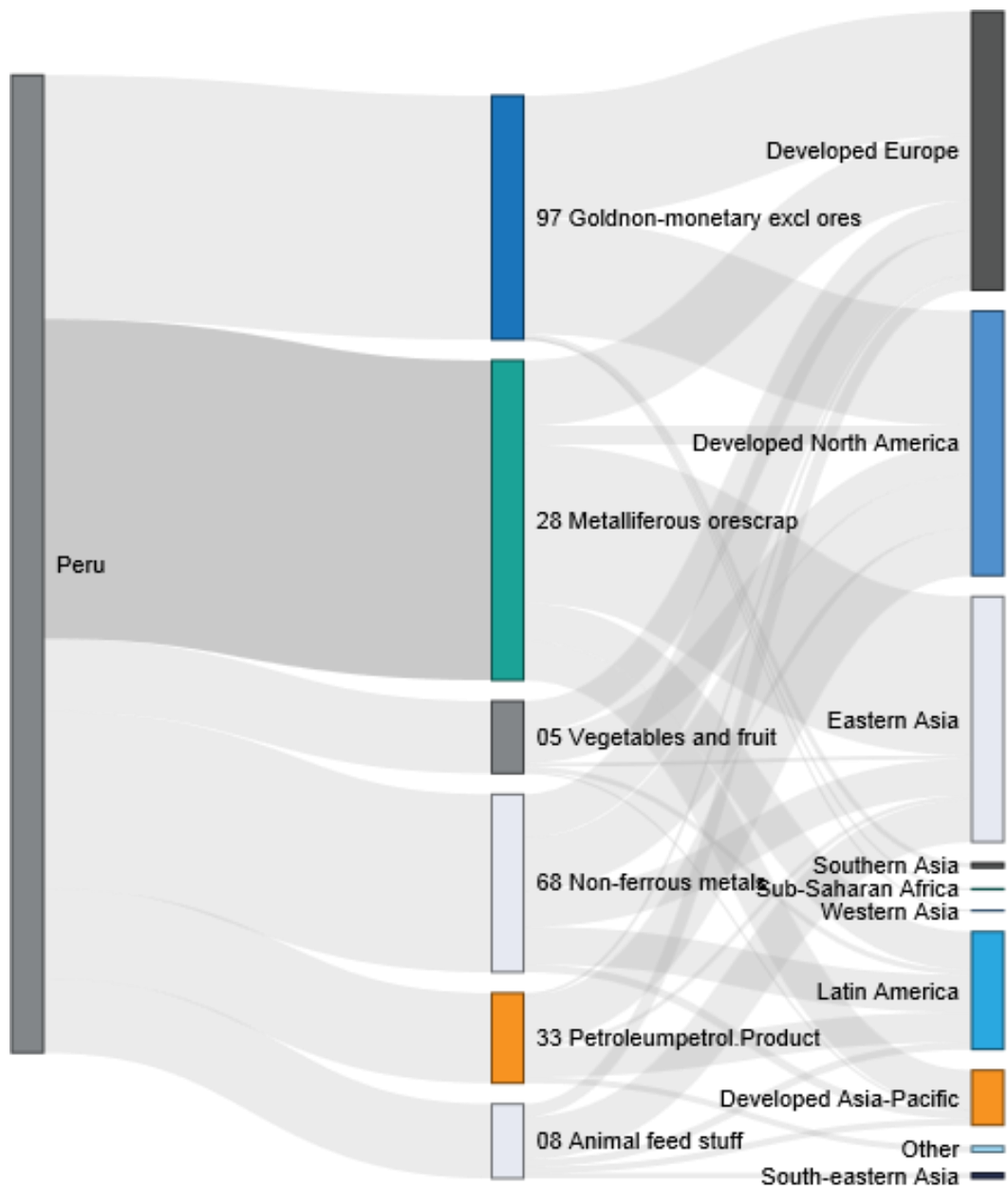

*Graph 1.1: This graph shows the destination and product type of Peru's exports flowing from the left to the right. The number in front of each type of goods represents the commodity code. The height of each rectangle stands for the relative trade volume of the good. Source: UN Comtrade, Version 2016 Aug.*

## 2. Review of Gravity Model Applications

The gravity model for panel data has been frequently used in analyzing the export performance of a country, and less frequently in the performance of a specific trade industry within a broader economy. For example, a study was conducted by Deluna et al. [10] using OLS methods, to examine the factors influencing the transfer of goods between the Philippines and its trade partners, while applying the gravity model to panel data for the period 2008 to 2012. Geographical area of the country, membership in regional organizations and a group of indices related to the freedom of trade, investment levels, fiscal parameters and corruption has been incorporated into the standard gravity model of trade. The results show that China and members of Association of Southeast Asian Nations ("ASEAN") possessed the highest potential for merchandise export. A similar study, conducted by Karamuriro & Karukuza [11] analyzed the determining factors of exports in Uganda, covering the period from 1980 to 2012. Foreign exchange rate has been incorporated into the gravity model of trade in their study. The authors applied both generalized least square method and the instrumental variables Generalized Method of Moments ("GMM") regression. They concluded that the formation of the Common Market for Eastern and Southern Africa ("COMESA") and East African Community ("EAC") provide significant benefits for Uganda's export value and that the process of regional economic integration should be deepened.

Within the same framework, the gravity model of trade has also been frequently applied to specific products in analyzing interesting patterns and determining major factors making up a certain component of the trade flow. An analysis on the volume of tourists from over twenty countries to Malaysia was conducted by Kosnan & Ismail to evaluate tourism sector performance [12]. The gravity model turned out to be suitable for this case. The authors advised Malaysia to further develop ties with neighboring countries in close vicinity to Malaysia since these countries made more contribution to export value growth. Similarly, a study conducted by Tian & Yu analyzed Chinese imported fruits using the gravity model, incorporating an index of quality [13]. They used OLS, fixed effect regression, random effect regression and instrumental-variable fixed effects regression. They concluded that quantity is still under quality for importing fruits in China.

On the other hand, a research paper by Ruta & Venables with a large literature review pointed out that the major factors influencing international trade of natural resources can be distinctive from other commodities that were covered in most of the studies above on international trade. National interests such as increasing government revenue, reducing domestic prices for consumers and stimulating domestic downstream production are given the priority when dealing with international trade in natural resources. Therefore, specific policies and inefficient long run contracts which were discussed through WTO forums and Doha negotiation [14].

Following previous gravity model studies, this report seeks to answer the following questions: What trends does the gravity model reveal within the international trade flows of Peru? Is the same model useful in analyzing a specific component of the export system (in this case copper)? What interesting and useful conclusions can be drawn from our results? For example, is the membership in MERCOSUR beneficial to Peru?

We call our export analysis of the general economy model of Peru (GMP), our analysis of the copper trade of Peru (CTP), and our regional trade agreement export analysis (RTP).

## 3. Methodology and Data

### 3.1. Conceptual Framework of the Gravity Model

The gravity model of trade was applied in this study considering the aforementioned power to explain real world data. Tinbergen [15] and Poyhonen [16] applied the gravity model for the first time to study trends in global trade in the 1960s. The gravity model's popularity as a useful instrument in the empirical analysis of foreign trade has continued to date. The gravity model of trade is defined most simply as:

$$X_{ij} = \frac{KY_i^{\alpha}Y_j^{\beta}}{D_{ij}^{\theta}} \qquad (3.1)$$

From Equation 3.1, $X_{ij}$ stands for the transaction volume between countries $i$ and $j$; $Y$ stands for the value of nominal GDP of the trading partners; $D_{ij}$ is the absolute physical distance between the trading partners; $K$ is a constant. Equation 3.1 can be converted into log-linear form as:

$$lnX_{ij} = lnK + \alpha lnY_i + \beta lnY_j - \theta lnD_{ij} + \delta Z + u_{ij} \qquad (3.2)$$

From Equation 3.2, the variable $\delta Z$ represents any hidden factors that could affect export performance, while $u_{ij}$ is the stochastic term. Following the concept of the gravity model in physics, given the same distance, larger economic mass will lead to stronger gravity force between these two objects – i.e. larger trading value. On the other hand, given a fixed economic mass product of the trading partners, the closer these two countries, the more frequent transactions of or higher volume of exports between them will take place. Following the traditional approach of the gravity model of trade, additional variables, such as population and sharing of boarders are added to better depict the transaction environment. Thus, we can write the augmented gravity model as:

$$X_{ij} = \beta_0 Y_i^{\beta_1} Y_j^{\beta_2} N_i^{\beta_3} N_j^{\beta_4} D_{ij}^{\beta_5} A_{ij}^{\beta_6} e^{ym} \varepsilon^{u_{ij}} \qquad (3.3)$$

From Equation 3.3, $X_{ij}$ is the value of exports between pairs of countries, $Y_i$ ($Y_j$) represents the GDP of the exporter (importer), $N_i$ ($N_j$) is the population of the exporter (importer), and as above $D_{ij}$ is the physical distance between the economic centers of the two countries, $A_{ij}$ represents other possible variables that could either hinder or ameliorate exports to another country.

### 3.2. The Model Applied In This Report

In its basic form, the hypothesis of the gravity model of bilateral trade states that "exports between two countries" are positively "related to their economic mass (measured by GDP and population) and inversely proportional to the distance." [17]. Empirical works such as Bergstrand [18] have provided several additions to the gravity model. An appropriate definition of the model for international trade is:

$$X_{ijt} = \beta_0 Y_{it}^{\beta_1} Y_{jt}^{\beta_2} N_{it}^{\beta_3} N_{jt}^{\beta_4} D_{ij}^{\beta_5} \varepsilon^{u_{ijt}} \qquad (3.4)$$

Following the empirical papers, latent but critical, additional variables should be incorporated into the gravity model with respect to geographical factors, national political inclinations and the general strategy of development of a country.

Dummy variables are added for the regional organizations, common border and common language under study. Thus, the augmented gravity model becomes;

$$X_{ijt} = \beta_0 Y_{it}^{\beta_1} Y_{jt}^{\beta_2} GDPPC_{it}^{\beta_3} GDPPC_{jt}^{\beta_4} GDPPCDIF_{ijt}^{\beta_5} D_{ij}^{\beta_6} FX_{ij}^{\beta_7} \beta_8 Language_{ij} \beta_9 APEC \\ \beta_{10} CAN \beta_{11} MERCOSUR \beta_{12} Border_{ij} \varepsilon^{u_{ijt}} \quad (3.5)$$

For more accessible equation solving, including in our analysis, the gravity model is commonly employed in its log-linear form. Hence, Equation 3.5 can be equivalently written using natural logarithms as:

$$lnX_{ijt} = ln\beta_0 + \beta_1 lnY_{it} + \beta_2 lnY_{jt} + \beta_3 lnGNIPC_{it} + \beta_4 lnGNIPC_{jt} \\ + \beta_5 lnGDPPCDIF_{ijt} + \beta_6 lnD_{ij} + \beta_7 FX_{ij} + \beta_8 Language_{ij} + \beta_9 APEC \\ + \beta_{10} CAN + \beta_{11} MERCOSUR + \beta_{12} Border_{ij} + U_{ijt} \quad (3.6)$$

From Equation 3.6, *GDPPCDIF$_{ijt}$* is the absolute value of the per capita GDP difference between countries *i* and *j* at time *t*, *FX* is the real exchange rate between countries *i* and *j* at time *t*. *Language$_{ij}$* equals to one if a country shares an official common language with Peru and zero otherwise, and *APEC* equals to one if a partner is an APEC member country and zero otherwise, *CAN* equals to one if a partner is a CAN member country and zero otherwise, *MERCOSUR* equals to one if a partner is a member or a corresponding member of MERCOSUR and zero otherwise. *Border$_{ij}$* equals to one if a common border exists between two trading partners and zero otherwise while *U$_{ijt}$* is a stochastic error term. This choice of variables follows the method of the Uganda paper [11]. In addition, in this paper, *GNIPC* took the place of *GDPPC*. GNI is the cumulative value of all resident producers combined plus any product taxes (but subtract subsidies) not included in the valuation of output plus net primary income (meaning compensation of employees and property income) from abroad. Since Peru is heavily involved with foreign trade, such as the copper trade in China, and would have many high income nationals working abroad, *GNIPC* is clearly more appropriate for this country.

Specific modification is needed when doing the estimation of the copper export of Peru, following the conclusion of Ruta & Venables [14]. Considering the fact that certain importers, such as China and countries in the European Union, account for significant proportions of Peru's trade value, the importer's demand reflected by GDP is expected to contain outliers and should be described using dummy variables for each outlier, showing the abnormally high domestic demand of the importers. In the view of Ruta & Venables' [14], international trade in the field of natural resources is subject to a range of government interventions. On the other hand, in the Keynesian view, aggregate demand does not necessarily equal the productive capacity of the economy; instead, it is influenced by a host of factors and sometimes behaves erratically: affecting production, employment, and inflation [19]. Therefore, the domestic inflation rate of the importers has been incorporated into the model of copper trade as a supplemental variable depicting the national demand. The value one has been added to the inflation/deflation rate of each year. Considering that the impact of domestic construction will have a lag before the investment turns into demand, a lag of 2 has been set into the *INFLATION* variable according to traditional lag analysis. Copper is an industry product whose demand is less related to consuming power of individuals. Therefore, *GNIPC* has been omitted when doing estimation on copper industry. Hence, Equation 3.6 should be modified as follows:

$$logX_{ijt} = log\beta_0 + \beta_1 logY_{it} + \beta_2 logY_{jt} + \beta_3 logD_{ij} + \beta_4 FX_{ij} + \beta_5 Language_{ij} + \\ \beta_6 Border_{ij} + \beta_7 \text{IND} + \beta_8 \text{KOR} + \beta_9 \text{CHL} + \beta_{10} \text{CHN} + \beta_{11} \text{USA} + \\ \beta_{12} \text{EU} + \beta_{13} \text{JPN} + \beta_{14} logIFL_{j(t-2)} + U_{ijt} \quad (3.7)$$

From Equation 3.7, *IND, KOR, CHL, CHN, USA, EU* and *JPN* are dummies that take value one if the country is India, Korea, Chile, China, United States, a member country of European Union and Japan respectively. *IFL* stands for the annual inflation rate of country *j* with a lag of 2 years.

Below is an explanation of how each of the above factors was expected to affect Peru's exports:

(1) *GDP* in the model stands for the factors associated with the level of economic development [20]. The capacity to produce or manufacture of the exporting side and the purchasing ability of the importers can also be generally reflected through this variable as well. The higher the GDP is, the greater the potential supply and demand can be. Thus, the GDP variables' coefficients were expected to be positive.

2) The efficiency in communication can facilitate trade flows between countries. Language barriers between countries are expected to cause obstacles in business communication and therefore reduce the chance of trading. Therefore, a positive sign is predicted for the estimated coefficient for this variable.

3) The formation of regional economic organizations is anticipated to promote export volumes within a specific region. Therefore, the estimated coefficients of these variables were expected to have a positive sign or there would be no incentive for the country to remain in the agreement. A positive value would imply that the membership of APEC, CAN and MERCOSUR increased export value, and vice versa.

4) The distance variable stands for the physical distance between the locations of economic hubs within the trading partner countries. The transportation costs go up along with the increase in distance. An increase in transportation costs will always raise the unit price of the final product for sale, thus reducing its demand. Therefore, a negative effect on exports for this variable is expected.

5) GNI per capita was substituted for population in our model as has been the case in many previous studies, for instance, the Uganda paper [11]. GNI per capita of a country could suggest strong economic self-sufficiency and less need for trade, or, the opposite, could stand for a stronger desire to consume and thus a drive to trade in a larger variety of goods. The coefficients of the GNI per capita were thus indeterminate.

6) The absolute difference in per capita *GDP, GDPPCDIF$_{ijt}$* has been added to the model to capture technological inequalities between countries engaged in trade. This variable is specifically used as a test of the Linder Hypothesis, which posits that "countries with similar levels of income per capita will exhibit similar behavior, produce similar but differentiated products and trade more amongst themselves." A negative sign on the per capita GDP difference variable would provide the necessary support for this hypothesis.

7) Our model utilized the real exchange rate ("*REAL*") as a proxy for relative prices. Currency appreciation makes a country more costly for others in foreign markets and therefore can reduce the competitiveness in price. We predicted the coefficient of the real exchange rate would be negative, implying that an appreciation hinders trade.

8) Inflation ("*IFL*") has been introduced in this study as a proxy for national demand in copper industry as discussed above. Investment in citizen education, transportation, real estate, and even health care can expand an economy in larger amounts than the original investment spending. The coefficient of the inflation was thus predicted to be positive if a growth in demand helped export value.

*3.3. Data Type and Sources*

The GMP used annual panel data on Peru and its trading partners for the period 2006 to 2015, including 108 different countries. The CTP study employed annual panel data on Peru and its trading partners for the period 2006 to 2015, including 21 different countries. The RTP study used annual panel data on Peru and its trading partners for the period 1994 to 2015 including only formal members or associate members of MERCOSUR. All data was obtained from UN Comtrade's datasets.

We used the amount of US dollars from Peru to each target export country as the dependent variable of the study. The data on exports was taken from the IMF *Direction of Trade Statistics* and the UN *Commodity Trade Statistics* (UN Comtrade) databases. The data on GDP, per capita income in USA dollars, the exchange rates and inflation rate were accessed using the *World Development Indicators* databases of the World Bank. Distance in kilometers and common language were obtained from *http://www.cepii.fr/CEPII/En/bdd_modele/ (2016)*.

## 4. Empirical Results and Discussion

*4.1. Diagnostic Tests*

Before fitting the coefficients of Equation 3.6 and 3.7, we did tests on the univariate characteristics of our datasets. First we explored the cointegrated relationship between the variables using a unit root test, which would be necessary if the variables are nonstationary. For the panel unit root tests, we choose the Im, Pesaran and Shin test ("IPS") and Fisher type test ("ADF Fisher test") [21].

We chose the Fisher and IPS tests since heterogeneity between the cross-section units is allowed in these tests, and simultaneous stationary and non-stationary data series are also allowed in these tests. The Fisher-type test does not require the panel to be balanced which is a further advantage. Test results are presented in Table A4.1.1 in the Appendix, showing that all variables are stationary (meaning null of unit root is rejected), so ordinary regression can be used to fit Eq. 3.6 and 3.7.

*4.2. Estimation Procedure*

*4.2.1 General model of Peru (GMP)*

Considering the situation where the assumptions of homoscedasticity and autocorrelation can seldom be met in a real world analysis, in this case the common error component over individuals induces correlation across the composite error terms, making OLS estimation inefficient. This requires using some type of feasible generalized least squares (GLS) estimators [22]. Therefore, GLS regression was applied both under fixed and random effects. The Hausman test was then applied to check which result was more efficient. The null hypothesis supports the random-effects result. (see results in Appendix A4.2.1).

The Hausman test statistic rejected the null hypothesis, suggesting that the fixed effects (within) regression was significantly different from the result of random effects and therefore is more efficient. However, coefficients of time-invariant variables were dropped with a fixed effects model since the effects of the omitted variables was absorbed into the intercept term of the regression [23]. Therefore, we used the instrumental variables Generalized Method of Moments (IV-GMM) regression model. The IV-GMM is widely employed in panel models as it investigates the endogeneity problem, and estimates accurately in the presence of heteroskedasticity [24].

*4.2.2 Specific model of Peru's copper industry (CTP)*

Following the general model of Peru, GLS regression was applied both under fixed and random effects to equation 3.7. The Hausman test was also applied (see results in Appendix A4.2.2). The Hausman test results shows that the null hypothesis cannot be rejected, suggesting that the random effects regression was as efficient as the fixed effects. Therefore, GMM was not applied in CTP.

*4.2.3 Model of impact of MERCOSUR on Peru's exports (RTP)*

Following the GMP approach for the RTP study, GLS regression was applied both under fixed and random effects to Equation 3.6. The Hausman test was also applied (see results in Appendix A4.2.3). The Hausman test results rejected the null hypothesis, suggesting that the fixed effects (within) regression was more efficient. Following the same logic, GMM was applied accordingly.

*4.3. Estimation Results and Discussion*

Table 4.1 summarizes the empirical results of GMP obtained from estimating Equation 3.6, using fixed effects GLS regression, random effects GLS regression and instrumental variables GMM regression.

Table 4.2 summarizes the empirical results of the CTP study obtained from estimating Equation 3.7, using fixed effects GLS regression and random effects GLS regression.

Table 4.3 summarizes the empirical results of analysis of RTP obtained from estimating Equation 3.6, using fixed effects GLS regression, random effects GLS regression and instrumental variables GMM regression. Considering this analysis focuses on the impact of the membership in MERCOSUR, GNIPC was omitted during the estimation.

*4.3.1 Discussion on Peru's General Economy (GMP)*

As shown in Table 4.1, the effect of GDP of the importer and exporter were found to be positive and statistically significant, which is consistent with the prediction of the gravity model. The result implies that trading partners with a higher GDP demonstrate a higher demand and more chances to import and that an increase in the production capacity from Peru's side will trigger additional trading volumes. This is consistent with the findings of Carrillo and Lee [25] in their study of the effect of regional integration on both intraregional and intra-industrial trade in Latin America in the period 1980-1997.

The effect of GDP per capita difference was found to be negative and statistically significant in fixed effects and random effects estimated models. Its negative sign suggests that bilateral trade flows between Peru and its trading partners are related negatively to inter-country differences in the level of technological advancement. Therefore, the Linder hypothesis is supported. This suggests that Peru's domestic demand structure is similar to Peru's trading partners. The similar pattern has been recognized and discussed between the 'Four Tigers' East Asian New Industrial Countries by Chow, Kellman & Shachmurove [26].

The effect of official common language and border was found to be positive and statistically insignificant in all estimated models, which is not consistent with other countries' studies, but not unexpected in this situation. The likely interpretation to this result is that the official language of Peru, Spanish, is most commonly used in Latin America, but Spanish speaking nations account for only a small proportion of Peru's export flow relative to non-Spanish speaking nations so they are biased against in the model. In terms of the GMP, the coefficients were insignificant.

The regression results show that the effect of Peru' per capita income was negative and statistically significant at the 5 percent level in fixed effects and random effects estimated models. This implies an increase in Peru's per capita income raises the domestic demand, which can mostly be satisfied with domestic supply, resulting into lower exports. This also happens in the study of Uganda [11].

The effect of geographical distance was found to be negative and statistically significant in all estimated models, which is consistent with the theoretical expectation. Transportation cost is one of the critical factors that determines the performance of Peru's export.

The effect of APEC on Peru's exports was found to be positive and statistically significant in all estimated models. This result suggests that Peru's participation in APEC has increased the opportunity to trade with APEC member countries rather than with non-members. This result is consistent with a paper conducted by Dee et al [27], published in 1996. The paper suggested that APEC members eventually won huge real income gains over what the real income could otherwise have been without APEC.

Naively, the effect of MERCOSUR was also found to be negative and statistically significant in GMM regression result for the general model. At the first glance, this result implies that Peru's participation in MERCOSUR adversely affected the export volume. Such implication has been verified to be invalid in discussion on Table 4.3 when we studied only MERCOSUR countries and time effects to see whether Peru joining MERCOSUR since 2003 improved the export value. The GMP panel model results for this coefficient were inaccurate because members of MERCOSUR were trading at a scale which is much lower than the dominant trade partners such as US and China, which are not MERCOSUR members. Thus the panel data model associates non-MERCOSUR status with higher export value. In reality, MERCOSUR increased the export value if only the MERCOSUR members are studied as in our RTP study. The Philippines panel model study [10] did not need to do this closer analysis of trade agreement members, as the trade partners with the trade agreement had more balanced export values with non-members.

*4.3.2 Discussion on Peru's Copper Trade (CTP)*

The effect of GDP of the exporter was found to be positive and statistically significant in random effects models according to Table 4.2. This result suggests that in terms of copper industry, Peru's GDP is still a key determinant of the country's capacity to export. A higher GDP represents a higher production capacity of copper. This may also imply that the purchasing power on the worldwide demand side exceeds the production capacity of copper in Peru. This is consistent with a research report from Freedonia Group published in May 2015 [28]. The paper predicted a 4.2% annual growth in worldwide copper demand and a 3.7% annual growth in worldwide copper production.

The effect of inflation of importer with a lag of 2 was found to be positive and statistically significant in all results.

The effect of common border was found to be negative and statistically significant in random effects models. The result is inconsistent with the expectation of the general theory. One possible explanation is that given the special property of copper as a natural resource, political benefits or national interest have been considered foremost, which can be supported by the paper of Ruta & Venables [14].

The remaining coefficients mainly stand for the countries which played a significant role in Peru's copper export. *KOR, CHN and JPN* have positive and statistically significant coefficient while *USA* has negative and statistically significant coefficient. The results mean that in the sector of copper trade, Korea, China and Japan imported abnormally high amount of copper while USA purchased abnormally small amount of copper from Peru when compared with other trading partners. This indicates that the demand of the importers differs significantly between different countries and such effect will possibly overshadow the explanatory power of other variables used in this regression. More generally, this suggests that in some cases when applying gravity model of trade to a specific product or natural resource that is highly demanded by a relatively small amount of trade partners, the economic mass cannot sufficiently depict the demand of a trade partner. Additional variables should be incorporated.

*4.3.3 Discussion on Peru's Regional Trade Agreements (RTP)*

In RTP, the effect of GDP of exporter and importers, distance and APEC is consistent with the general analysis in 4.3.1.

The effect of GDP per capita difference between countries was found to be positive and statistically significant in GMM estimated models. Its positive sign suggests that bilateral trade flows between Peru and its trading partners are related positively to inter-country differences in the level of technological advancement. Therefore, the Heckscher–Ohlin hypothesis was supported. This may suggests the product category remained generally heterogeneous in the case of Peru when trading with MERCOSUR member or associate member countries.

The effect of official common language and border was found to be positive and statistically significant in all estimated RTP models, which is consistent with the predicted theory and not consistent with the result in GMP models. Such result can in some way verify the explanation given in 4.3.1 that the positive effect of common language and common border was masked by in the presence of the massive exporting going to non-Spanish speaking countries in the general model. As stated in the study on Uganda [11], a country with more balanced trade partners compared with Peru, the coefficient of language and border turns out to be significantly positive.

The effect of MERCOSUR and CAN was found to be positive and statistically significant in all RTP regression results. The result corrects the bias found in Table 4.3.1 and is consistent with the prediction of the theory. In this case, the respective discussion in 4.3.1 is supported as the benefit of the membership in MERCOSUR was blinded by other factors in the general model. The membership in both MERCOSUR and CAN actually improve the trade between Peru and other MERCOSUR member or associate member countries.

## 5. Conclusion and Policy Implications

*5.1. Conclusion*

This study has examined the major determining factors in Peru's export value. An augmented trade gravity model was estimated using fixed effects GLS regression, random effects GLS regression and instrumental variables GMM regression. The study was separated into three aspects: general performance of Peru's export (GMP), performance of Peru's copper exports (CTP) and the impact of the membership in MERCOSUR (RTP).

For the GMP, the results from the instrumental variables GMM model showed that Peru's GDP, importer's GDP, Peru's per capita income, distance reduction and participation in APEC had a positive and statistically significant effect on Peru's exports. However, the substantial preference to export to Asian, North American, and European countries meant further, more zoomed-in, analysis needed to be done with specific industries as in CTP and the coefficients related to regional trade agreements in RTP.

For the CTP, the results from the random effect GLS showed that the economic effect of Peru's GDP is consistent with the GMP. The negative effect of the common border may suggest geographical features' impact on export value can be undetermined in case of certain goods which are related to natural resources. We demonstrated that in countries with a few very dominating trade partners, the GDP of the importer can be an insufficient explanatory variable to depict the demand. In this case, additional variables should be introduced as supplement. In this paper, abnormally high/low demand was depicted by assigning each country with a dummy variable. As also mentioned in a study by Babri et al [29], at a given time point, trade flows may also be affected by long-term contracts or cultural/national trading tendencies. Therefore, they extended the traditional gravity model such that a fixed amount is separated from observed trade flows and residuals are subject to discrete choice [29].

For the RTP, the result from the GMM model showed that when analyzing Peru's export performance in a rather micro way, taking into account only the trade partners who are members of MERCOSUR, certain effects covered by the general GMP analysis in the first aspect can be revealed. Peru's participation in MERCOSUR was therefore found to have positive effect on its exports. This conclusion is also applicable to the effect of common language and common border. The effect of Peru's per capita income in this third aspect suggests a different pattern of trading goods when trading with MERCOSUR countries compared to the rest of the world.

*5.2. Policy Implications*

The study highlights the factors that influence Peru's exports. The factors that have a positive effect on Peru's exports should be promoted. The study shows that in terms of the highest level of examining exports from Peru in the GMP study, a growth in GDP will definitely promote its trading value. On the other hand, emphasis should be given to trade partners with higher GDP and higher GDP growth. The negative effect of distance implies that the transport cost and border or adjacency variables support further investment in transport and communications infrastructure to reduce the cost of shipping and international business. This would potentially have a major impact on Peru's exports. The active participation in APEC will continue to promote Peru's export industries. The membership in MERCOSUR has a positive effect in Peru's foreign trade as shown in the RTP analysis. Therefore, active participation in MERCOSUR is also advised.

In terms of the copper export in CTP, the result implies that further developing the production of copper can be vital to the performance on this export. Foreign direct investment in copper mining industry may be a sound tool to spur the general economic growth. Major importers like China, Korea and Japan have significant impact on that industry. Therefore, to avoid instability in Peru's economy, long term strategic co-operation with other trade partners should be considered and soundly maintained due to Peru's current strong dependence on copper from just a few dominant trade partners.

| Variables | Fixed effects | | Random effects | | GMM | |
|---|---|---|---|---|---|---|
| Peru's GDP | 2.082 (0.564) | *** | 1.778 (0.546) | *** | 0.476 (0.165) | *** |
| Importer's GDP | 0.597 (0.292) | ** | 1.053 (0.102) | *** | 1.169 (0.092) | *** |
| Peru's per capita income | -1.597 (0.581) | *** | -1.530 (0.565) | *** | -0.352 (0.240) | |
| Importer's per capita income | -0.205 (0.324) | | -0.208 (0.130) | | -0.115 (0.163) | |
| GDP per capita difference | -0.097 (0.048) | ** | -0.096 (0.046) | ** | -0.010 (0.118) | |
| Real exchange rate | -0.064 (0.054) | | -0.082 (0.043) | * | -0.035 (0.064) | |
| Distance | | | -2.005 (0.395) | *** | -2.376 (0.359) | *** |
| Common official language | | | 0.570 (0.683) | | 0.342 (0.543) | |
| Common border | | | 1.016 (1.789) | | 0.844 (0.993) | |
| APEC | | | 1.658 (0.498) | *** | 1.360 (0.471) | *** |
| CAN | | | 0.068 (1.322) | | 0.025 (0.647) | |
| MERCOSUR | | | -1.249 (1.224) | | -1.442 (0.660) | ** |
| Constant | -36.087 (9.456) | *** | -22.439 (9.848) | ** | | |
| R-squared | 45.26% | | 69.77% | | 70.28% | |
| Number of observations | 1080 | | 1080 | | 1080 | |
| Hausman test | 13.68 | ** | | | | |

Dependent variable: Exports

Standard errors in parentheses.

\*\*\*, \*\*, \*: statistically significant at 1%, 5% and 10% levels respectively.

*Table 4.1 GMP Results*

| Variables | Fixed effects | | Random effects | |
|---|---|---|---|---|
| Peru's GDP | 0.964 (1.559) | | 3.253 (0.931) | *** |
| Importer's GDP | 4.022 (1.942) | ** | 0.187 (0.297) | |
| Distance | | | -0.828 (1.538) | |
| Real exchange rate | 3.168 (2.698) | | -0.394 (0.292) | |
| Common official language | | | 0.004 (1.418) | |
| Common border | | | -4.602 (2.266) | ** |
| IND | | | 0.822 (1.743) | |
| KOR | | | 4.383 (1.935) | ** |
| CHL | | | 2.497 (3.227) | |
| CHN | | | 3.872 (1.789) | ** |
| USA | | | -11.668 (2.013) | *** |
| EU | -0.791 (2.117) | | 0.682 (0.979) | |
| JPN | | | 5.173 (1.826) | *** |
| IFL, lag=2 | 22.484 (10.766) | ** | 27.934 (10.583) | *** |
| Constant | | | -64.43 (29.359) | ** |
| R-squared | 13.90% | | 43.15% | |
| Number of observations | 124 | | 124 | |
| Hausman test | 1.6082 | | | |

Dependent variable: Copper Exports

Standard errors in parentheses.

\*\*\*, \*\*, \*: statistically significant at 1%, 5% and 10% levels respectively.

*Table 4.2 CTP Results*

| Variables | Fixed effects | | Random effects | | GMM | |
|---|---|---|---|---|---|---|
| Peru's GDP | 0.461 (0.193) | ** | 0.554 (0.114) | *** | 0.206 (0.086) | *** |
| Importer's GDP | 1.060 (0.179) | *** | 0.692 (0.048) | *** | 0.788 (0.111) | *** |
| GDP per capita difference | -0.112 (0.052) | ** | 0.103 (0.037) | *** | 0.124 (0.037) | *** |
| Real exchange rate | 0.003 (0.017) | | -0.114 (0.013) | *** | -0.101 (0.034) | *** |
| Distance | | | -0.426 (0.317) | | -1.268 (0.513) | ** |
| Common official language | | | 1.334 (0.394) | *** | 0.751 (0.289) | *** |
| Common border | | | 2.635 (0.402) | *** | 1.926 (0.341) | *** |
| APEC | | | 2.915 (0.324) | *** | 2.532 (0.173) | *** |
| CAN | 0.484 (0.257) | * | 0.546 (0.315) | * | 0.643 (0.173) | *** |
| MERCOSUR | 0.328 (0.094) | *** | 0.327 (0.109) | *** | 0.543 (0.232) | ** |
| Constant | -19.446 (2.197) | *** | -13.386 (3.686) | *** | | |
| R-squared | 55.81% | | 96.37% | | 95.86% | |
| Number of observations | 174 | | 174 | | 174 | |
| Hausman test | 39.1 | *** | | | | |

Dependent variable: Exports

Standard errors in parentheses.

\*\*\*, \*\*, \*: statistically significant at 1%, 5% and 10% levels respectively.

*Table 4.3 RTP Results*

| | GMP | | CTP | | RTP | |
|---|---|---|---|---|---|---|
| **Variables** | **IPS** | **ADF-fisher** | **IPS** | **ADF-fisher** | **IPS** | **ADF-fisher** |
| GDP of importer | -13.4069(0.0000) | 719.024(0.0000) | -3.43302(0.0003) | 69.4047(0.0000) | -7.37151(0.0000) | 80.5783(0.0000) |
| GDP of exporter | -10.2936(0.0000) | 795.273(0.0000) | -3.13224(0.0009) | 73.6364(0.0000) | -2.74581(0.0030) | 31.3868(0.0120) |
| Tradevalue | -4.93122(0.0000) | 444.909(0.0000) | -3.59342(0.0002) | 58.1177(0.0001) | -11.1883(0.0000) | 134.899(0.0000) |
| GNI of importer | -3.03143(0.0012) | 364.695(0.0000) | | | | |
| GNI of exporter | -6.64113(0.0000) | 391.659(0.0000) | | | | |
| FX | -13.4416(0.0000) | 793.76(0.0000) | -7.65032(0.0000) | 120.561(0.0000) | -6.54963(0.0000) | 71.5146(0.0000) |
| Inflation rate | | | -2.41564(0.0079) | 61.5652(0.0000) | | |

*Appendix 4.1.1 Panel Unit Root Test Results*

|  | Coefficients | | | |
| --- | --- | --- | --- | --- |
|  | (b) | (B) | (b-B) | sqrt(diag(V_b-V_B)) |
|  | Fixed effect | Random effect | Difference | S.E. |
| lngdp | 0.5978124 | 1.053708 | -0.455896 | 0.2731703 |
| lnprgdp | 2.08268 | 1.778324 | 0.304356 | 0.1426165 |
| lngdppcdiff | -0.0973337 | -0.0963037 | -0.00103 | 0.0139143 |
| lnfx | -0.0647418 | -0.0829712 | 0.0182294 | 0.0329075 |
| lngnipc | -0.2056755 | -0.20844 | 0.0027645 | 0.2971402 |
| lnprgnipc | -1.597462 | -1.530646 | -0.066816 | 0.1368091 |

b = consistent under Ho and Ha; obtained from xtreg

B = inconsistent under Ha, efficient under Ho; obtained from xtreg

Test: Ho: difference in coefficients not systematic

chi2(6) = (b-B)'[(V_b-V_B)^(-1)](b-B)

$\quad$ = 13.68

Prob>chi2 = 0.0335
(V_b-V_B is not positive definite)

*Appendix 4.2.1. Hausman Test Results*

```
Hausman Test

data:  log(mdca$TradeValue) ~ log(mdca$GDP) + log(FX) +
    +lag(log(IFL), 2) + log(mdca$PRGDP) + log(mdca$Distance)
    + Language + border + IND + KOR + CHL + CHN + USA + EU + JPN
chisq = 1.6082, df = 5, p-value = 0.9003
alternative hypothesis: one model is inconsistent
```

*Appendix 4.2.2. Hausman Test Results*

|  | Coefficients | | | |
| --- | --- | --- | --- | --- |
|  | (b) | (B) | (b-B) | sqrt(diag(V_b-V_B)) |
|  | Fixed effects | Random effects | Difference | S.E. |
| lngdp | 1.059978 | 0.6915828 | 0.3683952 | 0.1729987 |
| lnprgdp | 0.4605499 | 0.5537584 | -0.0932085 | 0.1557461 |
| lngdppcdiff | -0.1121754 | 0.1032196 | -0.215395 | 0.0361521 |
| lnfx | 0.0034852 | -0.1138473 | 0.1173325 | 0.010335 |
| can | 0.4844439 | 0.5463401 | -0.0618962 | . |
| mercosur | 0.3284182 | 0.3270702 | 0.001348 | . |

b = consistent under Ho and Ha; obtained from xtreg

B = inconsistent under Ha, efficient under Ho; obtained from xtreg

Test: Ho: difference in coefficients not systematic

chi2(6) = (b-B)'[(V_b-V_B)^(-1)](b-B)

$\quad$ = 39.10

Prob>chi2 = 0.0000
(V_b-V_B is not positive definite)

*Appendix 4.2.3. Hausman Test Results*